\documentclass[aps,prd,reprint]{revtex4-1}
\usepackage{url}
\usepackage{textcase}
\usepackage{bm}
\usepackage{amsmath}
\usepackage{amsfonts}
\usepackage{amssymb}
\usepackage{hyperref}
\usepackage[table]{xcolor}
\usepackage{multirow}
\usepackage{bbm}
\usepackage{graphicx}
\usepackage{tensor}
\usepackage{subfigure}
\usepackage{array}
\usepackage[normalem]{ulem}
\usepackage{float}
\usepackage{fancyhdr}
\usepackage{tikz} 
\usepackage{tikz-feynman,contour}
\usepackage{contour}

\usepackage{pdfpages}
\makeatletter
\AtBeginDocument{\let\LS@rot\@undefined}
\makeatother

\usepackage{natbib}
\renewcommand{\Re}{\mathrm{Re}}
\newcommand{\cc}{}

\renewcommand{\vec}[1]{\boldsymbol{#1}}

\begin{document}

\title{The curious behaviour of the scale invariant $(2+1)$-dimensional Lifshitz scalar}
\date{\today}
\author{Daniel K.~Brattan}
\email{danny.brattan@gmail.com}
\affiliation{INFN, Sezione di Genova, via Dodecaneso 33, I-16146, Genova, Italy. \\ \& \\
Interdisciplinary Center for Theoretical Study, University of Science and Technology of China, 96 Jinzhai Road, Hefei, Anhui, 230026 PRC.}

\begin{abstract}
{\ We demonstrate the existence of an exactly marginal deformation, with derivative coupling, about the free theory of a $(2+1)$-dimensional charged, Lifshitz scalar with dynamic critical exponent $z=4$ and particle-hole asymmetry. We show that the other classically scale invariant interactions (consistent with translational and rotational invariance) break the scale symmetry at the quantum level and find a trace identity for the stress-energy-momentum tensor complex. We conjecture the existence of bound states of $(N+1)$-particles, as a manifestation of broken scale invariance, when we turn on an attractive, classically scale invariant, polynomial interaction in charged, scalar Lifshitz field theories with dynamic critical exponent $z=2N$, $n \in \mathbbm{N}$.}
\end{abstract}

\maketitle

{\ Symmetries, and their breaking, play a central role in the description of physical systems. Of particular utility has been scale invariance which has found applications in the renormalisation group \cite{WILSON197475}, critical phenomena \cite{henkel1999conformal} and string theory \cite{polchinski2005string}. This is a symmetry which manifests in the system having no internal scales. While it is generally easy to write down a classical system with scale invariance, it is non-trivial for the symmetry to survive quantisation. Instead one typically finds a quantum anomaly \cite{PhysRev.177.2426,Bell1969,PhysRevD.34.674,1993AmJPh..61..142H,COLEMAN1971552}. Theories without such anomalies have a special status.}

{\ When scale invariance does exist at the quantum level it is often elevated to a more constraining symmetry: conformal invariance \cite{DiFrancesco:639405}. Conformal invariance significantly simplifies any theory where it is present. It does this through the introduction of an additional set of conserved charges, the special conformal charges, which allow many exact results to be obtained such as analytic unitarity bounds. A class of theories that are generally scale but not conformally invariant are those with ``Lifshitz symmetry'' \cite{Alexandre:2011kr}. Such theories are invariant under a scaling $\{t , \vec{x}\} \rightarrow \{ \Lambda^{z} t,\Lambda \vec{x}\}$ where $z \neq 1,2$ is a positive real number called the dynamic critical exponent. This scaling symmetry can be found in nature at the finite temperature multicritical points of certain materials \cite{PhysRevLett.35.1678,PhysRevB.23.4615}, in strongly correlated electron systems \cite{PhysRevB.69.224415,PhysRevB.69.224416,Ardonne:2003wa} and heavy fermion metals \cite{2012PhRvL.109q6404R}. Lifshitz symmetry may also have applications in particle physics \cite{Alexandre:2011kr}, cosmology \cite{Mukohyama:2010xz} and quantum gravity \cite{PhysRevD.57.971,Kachru:2008yh,Horava:2009if,Horava:2009uw,Gies:2016con}.}

{\ There are surprisingly few examples of interacting Lifshitz quantum field theories that are known to be scale invariant. Many of the examples in the literature seem to rely on applying gauge-gravity dualities \cite{Kachru:2008yh,Taylor:2008tg} which suffer from the fact that the field content of the strongly coupled boundary theory is often inferred rather than precise. This paper seeks to redress this unsatisfactory situation by providing a simple example of an interacting Lifshitz field theory that is scale invariant at the quantum level. In particular we shall establish the perturbative Lifshitz scale invariance of the complex scalar field theory with action
  \begin{eqnarray}
    \label{Eq:ScaleInvariantAction}
    \mathcal{S} &=& \int d^{2}x dt \; \left( \frac{i}{2} \left( \Phi^{*} \partial_{t} \Phi - \partial_{t} \Phi^{*} \Phi \right) \right. \nonumber \\
    &\;& \left. \hphantom{\int d^{2}x dt \; \left( \right.} -  \frac{1}{2m} |\vec{\partial}^2 \Phi|^2  - \lambda |\Phi|^2 |\vec{\partial} \Phi|^2 \right) \; . \cc   
  \end{eqnarray}
The low energy modes of the free theory have a gapless quartic dispersion relation, $\varepsilon \sim \frac{1}{2m}\vec{k}^{4}$, as a consequence of the scaling symmetry. By tuning parameters such dispersion relations can approximately appear in ultracold atom gases \cite{PhysRevA.91.033404,2015PhRvA..91f3634R,2015NatCo...6E8012P,PhysRevB.96.085140} and excitons living in bilayer \cite{2016PhRvB..93w5110S}.}

{\ In section \ref{sec:classical} we discuss the most general classical field theory of a complex scalar with a single time derivative and $z=4$ Lifshitz scale invariance. We then proceed to quantise the theory and consider the scattering of particles in section \ref{sec:quantum}. We shall show that the 1PI four point function associated with $\lambda$ in \eqref{Eq:ScaleInvariantAction} respects scale invariance to arbitrary order in the loop expansion. Moreover we shall demonstrate that the other classically scale invariant interactions that one can add to \eqref{Eq:ScaleInvariantAction} necessarily break that invariance at the quantum level. This includes a sextic polynomial interaction which can lead to the appearance of a bound state in the spectrum. In the penultimate section we discuss the Ward identites and charges of our theory before concluding with a discussion of the ways that our results can be extended.}

\section{Classical field theory}
\label{sec:classical}

{\ We consider the theory of a complex scalar in $(2+1)$-dimensions with $z=4$ Lifshitz scale invariance, rotational and translational invariance. The most general action governing such a field, allowing only first derivatives in time, is
  \begin{eqnarray}
    \label{Eq:GenericAction}
    &\;& \int d^{2}x dt \; \left( \frac{i}{2} \left( \Phi^{*} \partial_{t} \Phi - \partial_{t} \Phi^{*} \Phi  \right)  -  \frac{1}{2m} |\vec{\partial}^2 \Phi|^2  \right. \nonumber \\
    &\;& \hphantom{\int d^{2}x dt \; \left( \right.} - \lambda |\Phi|^2 \|\vec{\partial} \Phi\|^2 - \frac{\tilde{\lambda}}{3!^2} |\Phi|^{6} \nonumber \\
    &\;& \left. \vphantom{\frac{i}{2} } \hphantom{\int d^{2}x dt \; \left( \right.} - \frac{|\Phi|^2}{2} \left( (\zeta \vec{\partial}^2 \Phi)^{*} \Phi + \zeta \Phi^{*} \vec{\partial}^2 \Phi \right)   \right) \; , \cc \qquad
  \end{eqnarray}
where we have dropped scale invariant interactions which differ by total derivatives from those displayed and $\left\| \vec{v} \right\|$ represents the vector norm of $\vec{v}$ as opposed to $| v |$ which is the modulus of a scalar quantity. We note that under $\{t,\vec{x}\} \rightarrow \{ \Lambda^{4} t, \Lambda \vec{x}\}$ the scalar field transforms as $\Phi \rightarrow \Lambda^{-1} \Phi$ so that the action is invariant; $\Lambda>0$ is a real number. The couplings $\lambda$ and $\tilde{\lambda}$ are real while $\zeta$ is complex \footnote{We note that the potential $2$-body interaction $|\Phi|^2 \epsilon^{ij} \partial_{i} \Phi^{*} \partial_{j} \Phi$ is a total derivative and thus we have not included it.}. If desired the parameter $m$ can be absorbed into the definition of non-relativistic units.}

{\ This system \eqref{Eq:GenericAction} should be compared to the well-known case of the complex Schr\"{o}dinger scalar in $(2+1)$-dimensions with the scale invariant polynomial interaction $|\Phi|^4$. That theory has an action of the form
  \begin{eqnarray}
    \label{Eq:SchrComparison}
    &\;& \int d^{2}x dt \; \left( \frac{i}{2} \left( \Phi^{*} \partial_{t} \Phi - \mathrm{c.c}  \right)  -  \frac{1}{2m} \|\vec{\partial} \Phi\|^2 
    - \frac{v}{4} |\Phi|^4 \right) \; , \cc \qquad
  \end{eqnarray}
with $v$ a real number. The model described by \eqref{Eq:SchrComparison} has been studied extensively \cite{Bergman:1991hf,Gomes:1996px}. Classically it manifests a particular kind of non-relativistic conformal invariance  which is broken at the quantum level by the polynomial interaction term. This breaking results in the appearance of a bound state of two particles in the spectrum \cite{Bergman:1991hf}. Our Lifshitz field theory \eqref{Eq:GenericAction} differs in a couple of important respects: there are more scale invariant interactions (some of which contain spatial derivatives), the Lifshitz field does not have boost invariance and there is no conserved special conformal charge. Consequently, unlike \eqref{Eq:SchrComparison}, we shall demonstrate that there is a scale invariant, interacting  version of \eqref{Eq:GenericAction} at the quantum level when all couplings but $\lambda$ are zero. Further, as we shall see, the polynomial interaction of \eqref{Eq:GenericAction} will lead to a bound state of three particles rather than two. Not all is rosy however as, unlike the Schr\"{o}dinger theory, the lack of special conformal charge means a loss of some of the power of the conformal algebra, including constraints on the structure of scattering amplitudes and analytic unitarity bounds on operator dimensions \cite{Hagen:1972pd,Henkel:1993sg,Mehen:1999nd,Nishida:2007pj}.}

\subsection{Symmetries}

{\ The spacetime symmetries of \eqref{Eq:GenericAction} include the scaling symmetry, translation invariance and spatial rotation invariance. Unlike a Galilean or relativistic theory, there is no boost invariance. Translational invariance yields the canonical stress-energy-momentum (SEM) tensor complex, $\{ T\indices{_{t}^{t}}, T\indices{_{i}^{t}} , T\indices{_{t}^{i}} , T\indices{_{i}^{j}} \} $, as a set of conserved currents
  \begin{eqnarray}
    \label{Eq:SEMconservation}
    \partial_{t} T\indices{_{t}^{t}} + \partial_{i} T\indices{_{t}^{i}} = 0 \; , \qquad \partial_{t} T\indices{_{j}^{t}} + \partial_{i} T\indices{_{j}^{i}} = 0 \; . \cc
  \end{eqnarray}
The canonical charge densities are:
  \begin{eqnarray}
    \label{Eq:H}
    T\indices{_{t}^{t}} &=& \mathcal{H} \; , \\
    \label{Eq:Pi}
    T\indices{_{i}^{t}} &=& \mathcal{P}_{i} = - \mathrm{Im}\left[ \Phi^{*} \partial_{i} \Phi \right] \; , 
  \end{eqnarray}
where $\mathcal{H}$ is the Hamiltonian density (see integrand of \eqref{Eq:Hamiltonian}) and $\mathcal{P}_{i}$ is the momentum density.  The expressions for the spatial components of the SEM tensor, $T\indices{_{t}^{i}}$ and $T\indices{_{i}^{j}}$, are more complicated than \eqref{Eq:H} and \eqref{Eq:Pi} so we have relegated them to the supplementary note 2. Importantly the stress piece, $T_{ij}$, is not symmetric in its indices unless we add improvement terms \cite{Callan:1970ze,Hoyos:2013qna}.}

{\ For a scale invariant theory the SEM tensor typically obeys a ``trace-relation''. For \eqref{Eq:GenericAction}, under the scaling transformation $\{t,\vec{x}\} \rightarrow \{ \Lambda^{4} t, \Lambda \vec{x}\}$, we find a conserved current, $\{S^{t},S^{i}\}$, of the form
  \begin{eqnarray}
    S^{t} &=& 4 t \mathcal{H} + x^{i} T\indices{_{i}^{t}} \; , \cc \\
    S^{i} &=& 4 t T\indices{_{t}^{i}} + x^{j} T\indices{_{j}^{i}} - K^{j}  \;, \cc
  \end{eqnarray}
where $K^{i}$ is an expression in the scalar fields called the ``virial current''. Using \eqref{Eq:SEMconservation} it is readily seen that conservation of the scaling current, $\partial_{t} S^{t} + \partial_{i} S^{i} = 0$, is equivalent to the trace relation 
  \begin{eqnarray}
    \label{Eq:tracerelation}
    4 \mathcal{H} + T\indices{_{i}^{i}} = \partial_{i} K^{i} \; . \cc
  \end{eqnarray}
Typically it is the case that one can find improvement terms that make the SEM tensor symmetric in the spatial indices and such that the virial current term in \eqref{Eq:tracerelation} vanishes. For our Lifshitz theory the resultant SEM tensor has a rather complicated form; moreover only the interaction with coupling $\lambda$ in \eqref{Eq:GenericAction} will be scale invariant at the quantum level. In this latter case it is possible to find the desired improvement term and the resultant expression is given in the supplementary note 2.}

{\ In addition to the spacetime symmetries our action \eqref{Eq:GenericAction} has an  internal symmetry, $U(1)$ particle number. The classical local charge density is given by
  \begin{eqnarray}
    \label{Eq:U(1)charge}
    \mathcal{J}_{U(1)}^{t} = \rho = | \Phi|^2 \; , 
  \end{eqnarray}
while the spatial part of the conserved current is
  \begin{eqnarray}
    \mathcal{J}_{U(1)}^{i} 
    &=& - \frac{i}{2m} \left[ \partial^{i}\partial^2 \Phi^{*} \Phi - \partial^{2} \Phi^{*} \partial^{i} \Phi - \mathrm{c.c.} \right] \nonumber \\
    &\;& + i \lambda \rho \left[ (\partial^{i} \Phi^{*}) \Phi - \Phi^{*} \partial^{i} \Phi \right] \nonumber \\
    &\;& - i \rho \left[ \zeta (\partial^{i} \Phi^{*}) \Phi - ( \zeta \Phi )^{*} \partial^{i} \Phi \right] \; . \cc
  \end{eqnarray}
Unlike non-derivative couplings, such as $\tilde{\lambda}$, the derivative couplings make an appearance in the charge current. We expect that these terms will have important consequences for the flow of charge in our theory.}

\subsection{Ground states}

{\ The classical equations of motion for the complex scalar field are
  \begin{eqnarray}
	\label{Eq:ClassicalEOM}
	i \partial_{t} \Phi
    &=& \frac{1}{2m} \vec{\partial}^{4} \Phi + \frac{\tilde{\lambda}}{12} \rho^2 \Phi - \lambda \rho \vec{\partial}^2 \Phi \nonumber \\
    &\;& - \lambda  \Phi^{*}  \vec{\partial} \Phi \cdot \vec{\partial} \Phi  + \zeta \rho \vec{\partial}^2 \Phi  \nonumber \\
    &\;&  + \zeta^{*} \left( \vec{\partial}^{2} \Phi^{*} \Phi^2 + 2 \Phi \| \vec{\partial} \Phi \|^2 
	  + \rho \vec{\partial}^2 \Phi \right. \nonumber \\
    &\;& \left. \hphantom{+ 2 \zeta^{*} \left( \right.} + \Phi^{*} \vec{\partial} \Phi \cdot \vec{\partial} \Phi \right)  \; , \cc \qquad
  \end{eqnarray}
and its complex conjugate. We shall be interested in perturbing about the vacuum state $\Phi=0$, by adding particles, and consequently we should check under what conditions it is the na\"{i}ve ground state of the system. To do this consider the Hamiltonian corresponding to \eqref{Eq:GenericAction} which is
  \begin{eqnarray}
    \label{Eq:Hamiltonian}
	H
    &=& \int d^{2}x \; \left[ \frac{1}{2m} \left|\vec{\partial}^2 \Phi + m \zeta \rho \Phi \right|^2  + \lambda \rho |\vec{\partial} \Phi|^2   \right. \nonumber \\
    &\;& \left. \vphantom{\frac{i}{2}} \hphantom{\int d^{2}x \; \left[ .\right.} + \frac{1}{3!^2} \left( \tilde{\lambda} - 18 m |\zeta|^2 \right) \rho^3 \right] \; .  \cc
  \end{eqnarray}
If we restrict ourselves to only considering states of positive or zero particle number, so that $\rho \geq 0$, then we can see $\Phi=0$ is certainly the minimal (zero) energy ground state if $\lambda\geq0$ and $\tilde{\lambda}\geq 18 m |\zeta|^2$. This follows from the classical Hamiltonian being manifestly positive under these constraints. As we shall see however, if $\tilde{\lambda}$ is non-zero the system will be unstable to forming a negative energy bound state of three particles indicating that $\Phi=0$ is no longer the minimal energy configuration when $\rho > 0$.}

\section{The quantum theory and scattering}
\label{sec:quantum}

{\ The classical free field solutions to the Lagrangian equations of motion \eqref{Eq:ClassicalEOM} have the form
  \begin{eqnarray}
   \Phi(t,\vec{x}) &=& \int \frac{d^{2} k}{(2\pi)^2} \; a(\vec{k}) e^{- i (\varepsilon t - \vec{k} \cdot \vec{x})} \; , \cc \\
   \Phi^{*}(t,\vec{x}) &=& \int \frac{d^{2} k}{(2\pi)^2} \; a^{*}(\vec{k}) e^{i (\varepsilon t - \vec{k} \cdot \vec{x})} \; , \cc
  \end{eqnarray}
where $\varepsilon = \vec{k}^4/(2m)$. We canonically quantise the theory by imposing the commutation relation
  \begin{eqnarray}
    \left[ \hat{\Phi}(t,\vec{x}) , \hat{\Phi}^{\dagger}(t,\vec{y}) \right] = \delta^{(2)}\left(\vec{x} - \vec{y}\right) \; , \cc
  \end{eqnarray}
and posit the existence of a free vacuum state such that $\hat{a}(\vec{k}) |0 \rangle=0$. Given the global $U(1)$ number symmetry we can interpret $\hat{a}(\vec{k})$, and subsequently $\hat{\Phi}(t,\vec{x})$, as an operator which destroys particles while $\hat{a}^{\dagger}(\vec{k})$ creates them.}

{\ The propagator of the free theory is
  \begin{eqnarray}
   \label{Eq:Freepropagator}
   D(r,t) &=& - i  \theta(t) \int_{0}^{\infty} \frac{d \| \vec{k} \|}{(2\pi)} \| \vec{k} \| J_{0}(\| \vec{k} \|r) e^{i \left( \frac{\vec{k}^4}{2m} t \right)} \; , \qquad
  \end{eqnarray}
where $J_{0}$ is the zeroth order Bessel function of the first kind and $r$ is the radius. While the integral over $\| \vec{k} \|$ can be performed explicitly, and the result is expressed in terms of generalised hypergeometrics, the expression is cumbersome and we shall not need its explicit form here. The presence of the step function in \eqref{Eq:Freepropagator} ensures that particles only propagate forwards in time. Moreover, global charge conservation implies that the number of scalar particles in any process cannot change. This significantly reduces the number of Feynman diagrams we need to consider in perturbation theory.}

\subsection{The exact propagator}

{\ The bare 1PI two point function will have the form
  \begin{eqnarray}
   \Gamma_{2}^{\mathrm{1PI}}(\vec{p},\varepsilon) = \varepsilon - \frac{1}{2m} \vec{p}^4 - \Sigma(\vec{p},\varepsilon) - \Pi(\vec{p},\varepsilon) \cc
  \end{eqnarray}
where $\Sigma$ represents self-energy contributions and $\Pi$ represents all other perturbative corrections. Typically for a scalar field theory with at least a quartic interaction, which includes our theory \eqref{Eq:GenericAction}, these corrections have a diagrammatic expansion similar to
  \begin{eqnarray}
     \label{Fig:Sigma}
      - i \Sigma &=& \begin{tikzpicture}[baseline=(a)]
      \begin{feynman}[inline=(t)]
      \vertex (a) ;
      \vertex [right= of a] (b);
      \vertex [above= of b] (t);
      \vertex [right= of b] (c);
  
      \diagram* {
	(a) -- [charged scalar] (b),
	(b) -- [charged scalar, half left] (t),
	(t) -- [charged scalar, half left] (b),
	(b) -- [charged scalar] (c),  
	};
    \end{feynman} 
   \end{tikzpicture} +
   \begin{tikzpicture}[baseline=(a)]
      \begin{feynman}[inline=(t)]
      \vertex (a) ;
      \vertex [right= of a] (b);
      \vertex [above= of b] (t1);
      \vertex [above= of t1] (t2);
      \vertex [right= of b] (c);
  
      \diagram* {
	(a) -- [charged scalar] (b),
	(b) -- [charged scalar, half left] (t1),
	(t1) -- [charged scalar, half left] (b),
	(t1) -- [charged scalar, half left] (t2),
	(t2) -- [charged scalar, half left] (t1),
	(b) -- [charged scalar] (c),  
	};
    \end{feynman} 
   \end{tikzpicture} + \ldots \; , \cc
  \end{eqnarray}
while $\Pi$ is given by
  \begin{eqnarray}
     \label{Fig:Pi}
      - i \Pi &=&  \begin{tikzpicture}[baseline=(a)]
      \begin{feynman}[inline=(a)]
      \vertex (a) ;
      \vertex [right= of a] (b);
      \vertex [right= of b] (c);
      \vertex [right= of c] (d);
  
      \diagram* {
	(a) -- [charged scalar] (b),
	(b) -- [charged scalar, half left] (c),
	(b) -- [charged scalar] (c),
	(b) -- [anti charged scalar, half right] (c),
	(c) -- [charged scalar] (d),
	};
	
    \end{feynman} 
   \end{tikzpicture} + \ldots \; . \cc
  \end{eqnarray}  
The latter term, $\Pi$, evaluates to zero for charged theories with a single time derivative. This is because there are no $1 \rightarrow 1$ loop diagrams which conserve charge and only contain particles travelling forwards in time. However, the self-energy contributions $\Sigma$ involve evaluations at the same spacetime point and can still contribute to the propagator in perturbation theory. If we use creation-annihilation operator normal ordering for our interactions these self-energy terms will also evaluate to zero. As the ordering should not affect results we assume this going forward.}

{\ Throughout this paper we shall use the continuum renormalization procedure; as opposed to Wilson's procedure for which a partial result for our field theory can be found in \cite{PhysRevA.91.033404}. Our action contains five bare parameters (the couplings $\lambda$, $\tilde{\lambda}$ and $\zeta$ plus field normalisation and the size of the gap which we have taken to be zero). Choosing as a renormalisation condition
  \begin{eqnarray}
    \label{Eq:Exact2pt}
    \Gamma_{2}^{\mathrm{1PI,R}}(\vec{p},\varepsilon) = \varepsilon - \frac{1}{2m} \vec{p}^4 \; , \cc
  \end{eqnarray}
where $R$ indicates the renormalised $2$-point function, consequently means that we do not require gap or field renormalisation counterterms. The exact propagator is the inverse of the 1PI two point function given in \eqref{Eq:Exact2pt}.}

\subsection{1PI four point functions}

\begin{figure}[!t]
  \centering
  \includegraphics[width=\columnwidth]{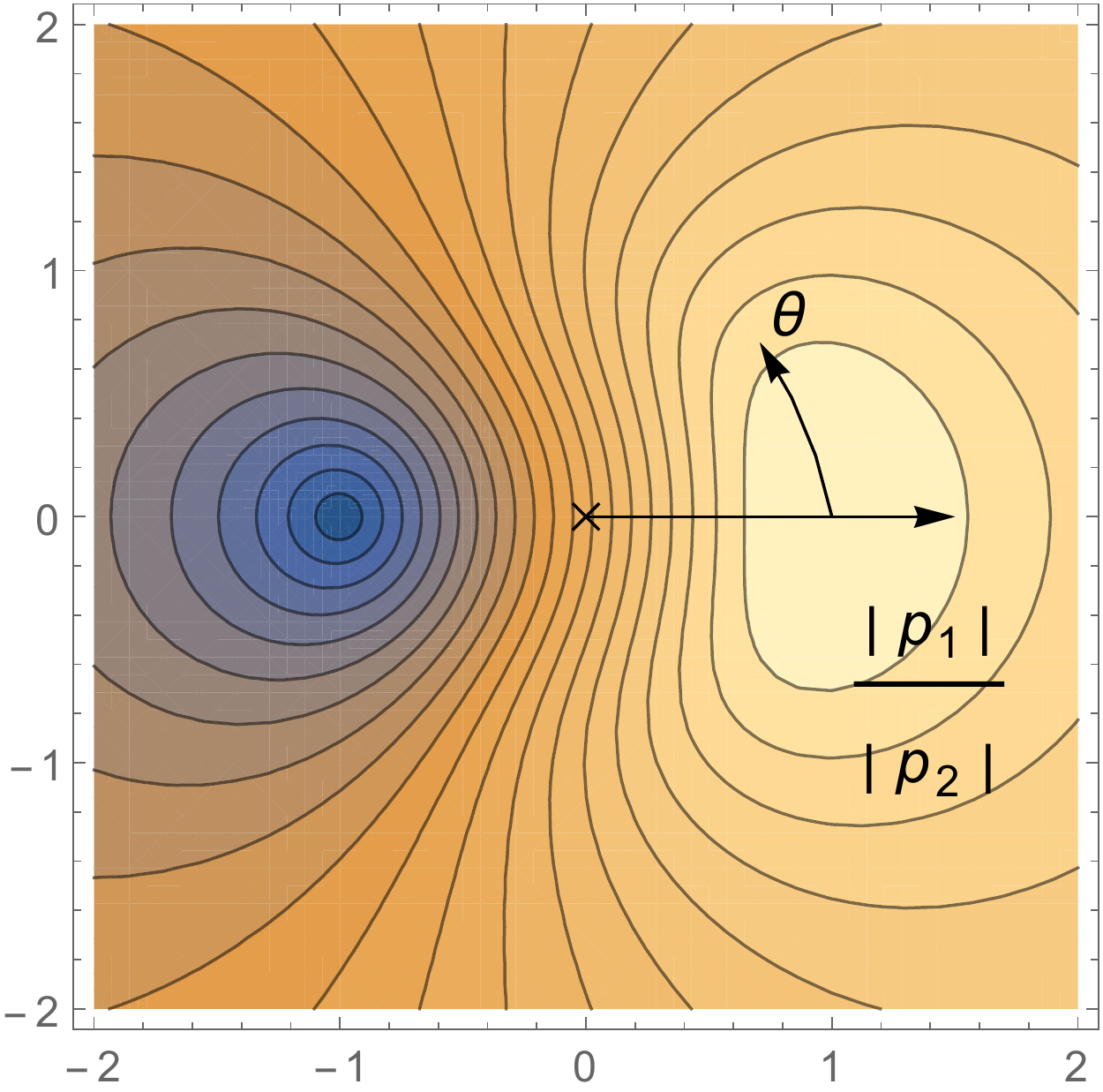}
  \caption{A contour plot of the on-shell kinematic parameter for two body scattering, $X=\left\| \vec{P} \right\|/(2 m \Omega)^{1/4}$, as a function of the ratio of the absolute value of the particle momenta, $p_{1}/p_{2}$, against the angle between the particles $\theta$. The minimum value of the parameter is zero (blue trough on left-hand side) and its maximum value is $2^{3/4}$ (yellow peak on right hand side). This parameter is conserved across $2$-body scattering vertices.}
  \label{Fig:X2onshell}
  \vspace{-1em}
\end{figure}

{\ For on-shell two body scattering in $(2+1)$-dimensions, in the absence of boost invariance, we require four independent variables to completely specify a scattering process. We can take for example: the smallest angles between incoming and outgoing particles, the total initial energy $\Omega$ and the two-body kinematic parameter
  \begin{eqnarray}
   \label{Eq:KinematicParameter}
   X = \frac{\left\| \vec{P} \right\|}{(2 m \Omega)^{\frac{1}{4}}} \cc
  \end{eqnarray}
where $\vec{P}$ is the total incoming/outgoing momentum. The angles and $X$ are scale invariants while $\Omega$ scales as $\Omega \rightarrow \Lambda^{-4} \Omega$ under $\{t,\vec{x}\} \rightarrow \{ \Lambda^{4} t, \Lambda \vec{x} \}$. }

\begin{figure}[!t]
  \centering
  \includegraphics[width=\columnwidth]{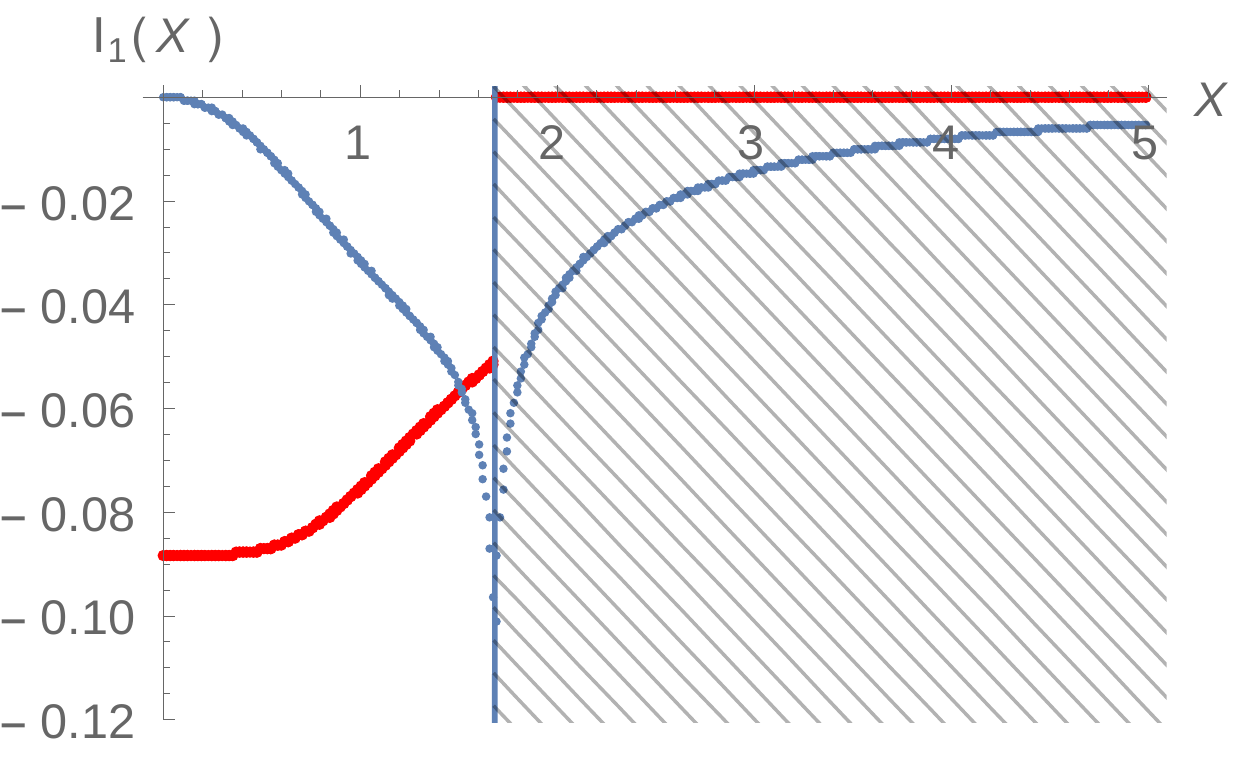}
  \caption{A plot of the real (blue dots forming a peak) and imaginary (red dots that are approximately zero in the shaded region) parts of the one-loop contribution \eqref{Eq:OneLoopCorrection} to the scattering amplitude from two $\lambda$ vertices. We plot the value of this contribution against the kinematic variable $X$ with $\epsilon=10^{-5}$. The shaded region has a kinematic parameter greater than the maximum on-shell value.}
  \label{Fig:Finite2BodyInteraction}
  \vspace{-1em}
\end{figure}

{\ On-shell the kinematic parameter $X$, given by \eqref{Eq:KinematicParameter}, belongs to the interval $[0,2^{3/4}]$ and we have plotted it as a function of the scattering angle between two incoming (outgoing) particles and the ratio of the magnitude of their momenta in fig.~\ref{Fig:X2onshell}.  We note that, just as for a theory with quadratic dispersion, $E=\vec{P}^2$, a single particle state with momentum $\vec{P}$ in a Lifshitz field theory can have greater energy than a two particle state with total momentum $\vec{P}$ (see \cite{Fujimori:2015mea}). For \eqref{Eq:Hamiltonian} this occurs when two body kinematic parameter, $X$, is greater than one. Particle number conservation in our model \eqref{Eq:GenericAction} prevents the one particle state decaying into two particles; which would not be the case for a theory with particle-hole symmetry (i.e.~one having an action which is second derivative in time). This further reduces the number of Feynman diagrams we need to consider.}

{\ To begin with we take the $\lambda$ interaction in isolation and compute the one-loop correction to the two-body interaction coming from two such vertices. This corresponds to evaluating the Feynman diagram
  \begin{eqnarray}
     \label{Fig:OneLoopCorrection}
       \begin{tikzpicture}[baseline=(a)]
      \begin{feynman}[inline=(a)]
      \vertex (a) ;
      \vertex [above left=2cm of a] (i1);
      \vertex [below left=2cm of a] (i2);
      \vertex [right=2cm of a] (b);
      \vertex [above right=2cm of b] (f1);
      \vertex [below right=2cm of b] (f2);
  
      \diagram* {
	(i1) -- [charged scalar] (a),
	(i2) -- [charged scalar] (a),
	(a) -- [charged scalar, momentum={\(\varepsilon,\vec{p}\)}, half left, looseness=1.5] (b),
	(b) -- [anti charged scalar, reversed momentum={\(\Omega-\varepsilon,\vec{P}-\vec{p}\)}, half left, looseness=1.5] (a),
	(b) -- [charged scalar] (f1),
	(b) -- [charged scalar] (f2),     
	};
	
      \vertex [left=0.5em of a] {\(- i \lambda \vec{P}^{2}\)};
      \vertex [right=0.5em of b] {\(- i \lambda \vec{P}^{2}\)};
    \end{feynman} 
   \end{tikzpicture} \; . 
  \end{eqnarray}
Notice the factors of $\vec{P}^2$ in the coupling which come from spatial derivatives in the interaction. Up to an overall prefactor this correction, which we denote by $I_{1}(X)$, is given by evaluating
  \begin{eqnarray}
    \label{Eq:OneLoopCorrection}
    \int^{\tilde{\Lambda}} \frac{dx d \theta}{(2\pi)^2}  \frac{x}{1 - (X^2 - 2 X x \cos(\theta) + x^2)^2 - x^4 + i \epsilon} \qquad
  \end{eqnarray}
where $x=\|\vec{p}\|/(2 m \Omega)^{1/4}$, $\tilde{\Lambda}=\Lambda/(2m\Omega)^{1/4}$ and $\Lambda$ is a large momentum cut-off.}

{\ Unfortunately, several of the standard tricks for evaluating Feynman integrals fail in the Lifshitz case \eqref{Eq:OneLoopCorrection}. In particular, we do not have boost invariance so we cannot move to the centre of mass frame. Moreover, the quartic denominator cannot be factorised in such a way that we can use a variable redefinition to remove its dependence on the angle $\theta$ (c.f.~particles with linear and quadratic dispersion relations).}

{\ While in general the integral \eqref{Eq:OneLoopCorrection} does not have a closed form, an analytic result can be found when the incoming particles have momenta of equal magnitude but opposite sign, so that the total incoming momentum is zero. The value of the integral is finite as we take the cut-off to infinity is
  \begin{eqnarray}
   \label{Eq:OneLoopCorrectionX_{2}=0}
   \lim_{\epsilon \rightarrow 0^{+}} \lim_{\Lambda \rightarrow \infty} I_{1}(0) = - \frac{i}{8 \sqrt{2}} \; . \cc
  \end{eqnarray}
To compute this interaction at non-zero values of $X$ we can complete the square on the $\cos(\theta)$ in the denominator of \eqref{Eq:OneLoopCorrection} and integrate over $\theta$ such that we are left with a remaining integral over $x$  (see supplementary note 1 for the integral over $\theta$). This we evaluate numerically. The result is displayed in fig.~\ref{Fig:Finite2BodyInteraction}. We can readily see that at small $X$ we find the analytic value of \eqref{Eq:OneLoopCorrectionX_{2}=0}.}

{\ If all other couplings are zero (i.e.~$\zeta, \tilde{\lambda} \equiv 0$) then the 1PI four point function can readily be expanded in terms of the one-loop correction, as a consequence of charge conservation, to give the exact expression
  \begin{eqnarray}
    \label{Eq:4ptfunc}
   - i \Gamma_{4}^{\mathrm{1PI}} &=& \frac{- i \lambda \vec{P}^2}{1 + 2 m \lambda X^2 I_{1}(X)} \; . \cc
  \end{eqnarray}
Up to an overall factor of $\vec{P}^2$ this result is scale invariant indicating that $\lambda$ is an exactly marginal deformation about the free theory. The Gaussian fixed point is therefore one point on a line of interacting, scale invariant fixed points living in its neighbourhood.}

{\ This result for the four point function \eqref{Eq:4ptfunc} demonstrates one of the most pleasing features of non-relativistic scale invariance in a theory without boosts. For two body scattering in a relativistic conformal theory the coupling is fixed to be a constant as one cannot construct a Lorentz invariant scalar out of the relativistic momentum. However, in non-relativistic theories, there are two dimensionful kinematic parameters: the energy and the spatial momentum. This allows one to define a scale invariant kinematic object, $X$, upon which the coupling can depend. When Galilean boosts are a symmetry it is always possible to move to a frame where the total momentum, or equivalently $X$, is zero and any dependence of the coupling on $X$ is given by a boost from the rest frame. For our Lifshitz theory however, this is not the case and the behaviour of the theory is much richer.}

{\ While not our primary aim it is not hard to see how this result for the derivative interaction, namely exact scale invariance, would extend to systems with greater levels of anisotropy. Consider the free Hamiltonian
  \begin{eqnarray}
     \label{Eq:GenericzH}
     \frac{1}{2m} \int d^{2}x \;  \partial_{i_{1}} \ldots \partial_{i_{N}} \Phi^{*} \partial^{i_{1}} \ldots \partial^{i_{N}} \Phi \; . \cc
  \end{eqnarray}
The system it describes has the dispersion relation $\omega \sim \vec{k}^{2N}$ and anisotropic scaling symmetry of the form $\{t,\vec{x}\} \rightarrow \{\Lambda^{2N} t, \Lambda \vec{x}\}$. If we add to this system a two-body interaction term whose momentum space vertex is given by $- i \lambda \vec{P}^{2(N-1)}$ then in analogy with the results above the system remains scale invariant at the quantum level as the resulting Feynman diagram is finite.}

{\ The other two-body interactions in \eqref{Eq:GenericAction} give logarithmically divergent contributions to the four point function. In particular, the divergent one-loop diagrams containing one $\lambda$ vertex and one $\xi^{(*)}$ vertex are proportional to the integral $I_{2}(X)$,
  \begin{eqnarray}
    \label{Eq:OneLoopCorrection2}
    \int^{\tilde{\Lambda}} \frac{dx d \theta}{(2\pi)^2}  \frac{x \left(X^2 - 2 X x \cos(\theta) + 2 x^2 \right)}{1 - (X^2 - 2 X x \cos(\theta) + x^2)^2 - x^4 + i \epsilon} \; ,  \qquad
  \end{eqnarray}
which can be seen to yield a logarithmic divergence when $X=0$,
  \begin{eqnarray}
    \lim_{\epsilon \rightarrow 0^{+}} I_{2}(0) \stackrel{\tilde{\Lambda} \gg 1}{=} - \frac{1}{8 \pi} \left( i \pi + \ln(2) + \ln \left( \frac{\Lambda^{4}}{2 m \Omega} \right) \right) \; . \qquad
  \end{eqnarray}
Similarly, the divergent one-loop diagram containing both $\xi$ and $\xi^{*}$ has a quadratic divergence and a logarithmic divergence. Subtracting the quadratic term leaves
  \begin{eqnarray}
    \label{Eq:OneLoopCorrection3}
    \int^{\tilde{\Lambda}} \frac{dx d \theta}{(2\pi)^2}  \frac{- X^2 \left( X - 2 x \cos(\theta) \right)^2}{1 - (X^2 - 2 X x \cos(\theta) + x^2)^2 - x^4 + i \epsilon} \; , \qquad
  \end{eqnarray}
which we denote by $I_{3}(X)$. In the small $X$ limit this latter integral \eqref{Eq:OneLoopCorrection3} behaves as
  \begin{eqnarray}
   \lim_{\epsilon \rightarrow 0^{+}} \left[ I_{3}(X)/X^2 \right]_{X=0} \stackrel{\tilde{\Lambda} \gg 1}{=} \frac{1}{4 \pi} \ln \left( \frac{\Lambda^4}{2 m \Omega} \right)
  \end{eqnarray}
Numerical results for both of these integrals, \eqref{Eq:OneLoopCorrection2} and \eqref{Eq:OneLoopCorrection3}, in a range of $X>0$ can be found in supplementary note 2.}

\begin{figure}[!t]
  \centering
  \includegraphics[width=\columnwidth]{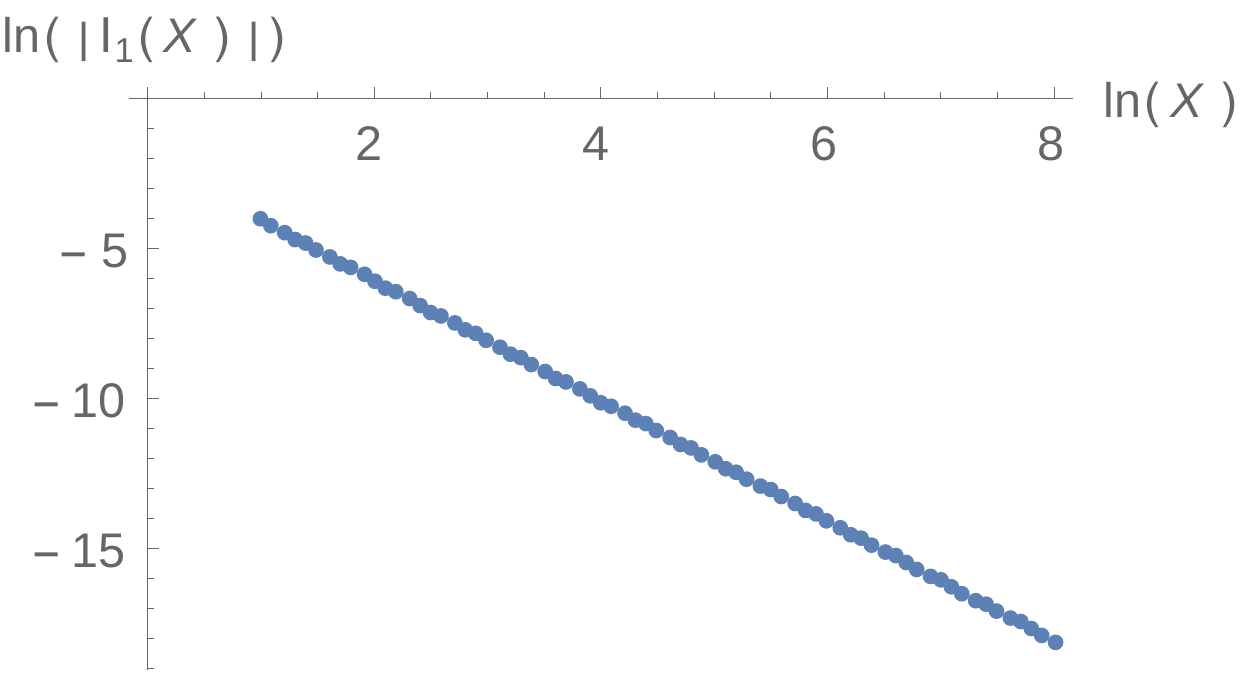}
  \caption{A plot of the logarithm of $\ln |I_{1}(X)|$ at large $X$ against $\ln|X|$. The behaviour at large $X$ is well approximated by a linear expression of the form $\ln|I_{1}(X)| = -4 \ln 2 - 2 \ln X$.}
  \label{Fig:largeXlimit}
  \vspace{-1em}
\end{figure}

{\ We can combine the above observations to determine the perturbative renormalisation group flow of our theory about the Gaussian fixed point. To define the $\beta$-functions we must pick a renormalisation scheme. An important ambiguity arises because the $2$-body operators associated with the three couplings $\lambda$, $\xi$ and $\xi^{*}$ can be mixed with each other at every loop order to define new operators and couplings. We can fix this ambiguity at one-loop by defining a minimal subtraction style scheme through the counterterm
  \begin{eqnarray}
    \begin{tikzpicture}[baseline=(a)]
    \begin{feynman}[inline=(a)]
      \vertex [crossed dot] (a) at (0,0) {\contour{white}{}};;
      \vertex [above left=of a] (i1) {\(\vec{k}_{1}\)};
      \vertex [below left=of a] (i2) {\(\vec{k}_{2}\)};
      \vertex [above right=of a] (f1) {\(\vec{k}_{3}\)};
      \vertex [below right=of a] (f2) {\(\vec{k}_{4}\)};
  
      \diagram* {
	(i1) -- [charged scalar] (a),
	(i2) -- [charged scalar] (a),
	(a) -- [charged scalar] (f1),
	(a) -- [charged scalar] (f2),
      };
    \end{feynman}
    \end{tikzpicture} &=& - \frac{i m}{\pi} \left( \frac{|\zeta|^2}{4} - \lambda \mathrm{Re}[\zeta] \right) \ln \left(\frac{\Lambda}{\mu}\right) \vec{P}^2 \nonumber \\
    &\;& - \frac{i m}{2 \pi} \Lambda^2 |\zeta|^2  \; , 
  \end{eqnarray}
where $\mu$ is some choice of energy scale. We notice that we have had to introduce a non-zero coupling for the IR relevant operator $|\Phi|^4$ because of the quadratic divergence coming from the divergent $\xi$-$\xi^{*}$ dependent diagram. This was to be expected when using momentum cut-off regularisation but we note that this counterterm will be zero if $\zeta=0$. With these new vertices we identify the $\beta$-functions at one-loop
  \begin{subequations}
  \begin{eqnarray}
    \label{Eq:lambda4flow}
    \mu d_{\mu} \lambda &=& \beta_{\lambda} = - \frac{m}{4 \pi} \left( \frac{| \zeta |^2}{4} - \lambda \mathrm{Re}[\zeta] \right) \; , \\
    \label{Eq:zeta4flow}
    \mu d_{\mu} \zeta^{(*)} &=& \beta_{\zeta^{(*)}} = 0 \; , 
  \end{eqnarray}
  \end{subequations}
where the displayed couplings are their renormalized values. These equations are valid in the vicinty of the free theory where we can solve them to find
  \begin{eqnarray}
	\lambda(\mu) 
    &=& \left\{ 
	  \begin{array}{ccc}
	    \left( \frac{\mu}{\mu_{0}} \right)^{\frac{m}{4 \pi} \mathrm{Re}[\zeta]} + \frac{|\zeta |^2}{4 \mathrm{Re}[\zeta]} & & \mathrm{Re}[\zeta] \neq 0 \\
	    - \frac{m \mathrm{Im}[\zeta]^2}{16 \pi} \ln \left( \frac{\mu}{\mu_{0}} \right) & & \mathrm{Re}[\zeta] = 0 
	  \end{array} \right. \; , \nonumber \\
    \zeta(\mu) &=& \zeta \; , \qquad
  \end{eqnarray}
with $\mu_{0}$ some initial choice of scale. We can see from this expression that relevance of the operator $|\hat{\Phi}|^2 \|\vec{\partial} \hat{\Phi}\|^2$ depends on the real part of the coupling $\zeta$. If this coupling is negative then we have an irrelevant operator in the IR, positive coupling leads to IR relevance and if $\zeta=0$ then it is marginal to first order in perturbation theory.}

{\ From \eqref{Eq:lambda4flow} and \eqref{Eq:zeta4flow} we see straightaway that if we move away from the free theory with $\zeta = 0$  then we remain on a fixed point of the renormalization group equations. There is seemingly a second fixed point given by setting the right hand side of \eqref{Eq:lambda4flow} to zero with arbitrary non-zero values of $\zeta$. Whether this is a true fixed point of the system will require a more delicate analysis.}

\subsection{1PI six point function}

{\ The leading correction to the 1PI six-point function requires the evaluation of a two loop diagram
  \begin{eqnarray}
     \label{Fig:TwoLoopCorrection}
       \begin{tikzpicture}[baseline=(a)]
      \begin{feynman}[inline=(a)]
      \vertex (a) ;
      \vertex [above left=2cm of a] (i1);
      \vertex [left=2cm of a] (i2);
      \vertex [below left=2cm of a] (i3);
      \vertex [right=2cm of a] (b);
      \vertex [above right=2cm of b] (f1);
      \vertex [right=2cm of b] (f2);
      \vertex [below right=2cm of b] (f3);
  
      \diagram* {
	(i1) -- [charged scalar] (a),
	(i2) -- [charged scalar] (a),
	(i3) -- [charged scalar] (a),
	(a) -- [charged scalar, half left, looseness=1.5] (b),
	(a) -- [charged scalar] (b),
	(b) -- [anti charged scalar, half left, looseness=1.5] (a),
	(b) -- [charged scalar] (f1),
	(b) -- [charged scalar] (f2),
	(b) -- [charged scalar] (f3),     
	};
	
    \end{feynman} 
   \end{tikzpicture} \; . 
  \end{eqnarray}
With the added difficulties posed by the lack of boost invariance evaluating the $6$-point function for arbitrary scattering processes is difficult (even numerically). Nonetheless we can determine its value readily in the case that the total incoming/outgoing momentum of the three scattered particles vanishes, and this is already sufficient to extract both the divergences and argue for the existence of a bound state. The existence of a bound state corresponding to the naively scale invariant polynomial interaction is in complete analogy with what is found in Schr\"{o}dinger field theory \eqref{Eq:SchrComparison}. Interestingly however, this bound state will be of three interacting particles.}

{\ We have found that the integral $I_{1}(X)$ behaves for large $X$ as
  \begin{eqnarray}
   I_{1}(X) \stackrel{X \gg 1}{\approx} - \frac{1}{8 X^{2}} \; . \cc 
  \end{eqnarray}
We determined this expression by numerically evaluating \eqref{Eq:OneLoopCorrection} up to $\ln X = 8$, as displayed in fig.~\ref{Fig:largeXlimit}, and fitting the large $X$ behaviour. Using this we can evaluate the lowest order correction to the six-point function which is proportional to the integral
   \begin{widetext}
    \begin{eqnarray}
    \label{Eq:OneLoopCorrection4}
    \int^{\tilde{\Lambda}} \frac{dx_{1} dx_{2} d \theta}{(2\pi)^3} \frac{x_{1} x_{2}}{1 - (x_{1}^2 + x_{2}^2 - 2 x_{1} x_{2} \cos(\theta) )^2 - x_{1}^4 - x_{2}^4 + i \epsilon} 
    &=& \int^{\tilde{\Lambda}} \frac{dx_{1}}{2 \pi} x_{1} I_{1}(x_{1}) \stackrel{\tilde{\Lambda} \gg 1}{=} \frac{1}{2 \pi} \left( \phi - \frac{1}{32} \ln \left( \frac{\Lambda^{4}}{2 m \Omega} \right) \right) \; , \cc \qquad
    \end{eqnarray}
  \end{widetext}
where $x_{1}=\|\vec{p}_{1}\|/(2 m \Omega)^{1/4}$, $x_{2}=\|\vec{p}_{2}\|/(2 m \Omega)^{1/4}$, $\vec{p}_{1}$, $\vec{p}_{2}$ are two of the loop momenta (the third is given by conservation) and $\theta$ is the angle between $\vec{p}_{1}$ and $\vec{p}_{2}$. The constant $\phi$ we determined numerically to be $\approx -0.307200 -0.098172 i$, such that the imaginary part is approximately $- i \pi /32$. Assuming this result is exact, the imaginary part of $\phi$ can subsequently be absorbed into the logarithm as an overall negative sign of the argument.}

{\ Simplifying our theory such that $\tilde{\lambda}$ is the only non-zero coupling we can expand the 1PI $6$-point function in terms of its leading correction, reminding ourselves of $3!$ symmetry factor and $(-i)^2$ from Cauchy integrals over two frequencies, to find
  \begin{eqnarray}
   \label{Eq:6pointfunction}
   \left. \Gamma_{6}^{\mathrm{1PI}} \right|_{\vec{P}=\vec{0}} &=& \frac{\tilde{\lambda}}{1 + \frac{m \tilde{\lambda}}{3! \pi} \left( \frac{1}{32} \ln \left( -\frac{\Lambda^{4}}{2 m \Omega} \right) - \Re[\phi] \right)}
  \end{eqnarray}
where $\vec{P}$ is the total ingoing/outgoing momentum. The pole in \eqref{Eq:6pointfunction} is indicative of a bound state with energy
  \begin{eqnarray}
    \Omega &=& - \frac{\Lambda^{4}}{2m} \exp \left( \left( \frac{192}{m \tilde{\lambda}} \right) \pi - 32 \Re[\phi] \right) \; . 
  \end{eqnarray}
}

{\ The existence of a bound state in the quantized theory is in complete analogy with the $|\Phi|^4$-Schr\"{o}dinger field \eqref{Eq:SchrComparison} with the exception that the bound state is now between three particles rather than two. A not dissimilar phenomenon happens for non-relativistic fermion fields with a sextic interaction in one dimension \cite{Drut:2018rip,Daza:2018nvg}. While we only have two data-points ($N=1,2$), our results hint that if we take the free Hamiltonian \eqref{Eq:GenericzH} and add an interaction term of the form
  \begin{eqnarray}
   V_{\mathrm{int.}} &=& \int d^{2}x \; \frac{\tilde{\lambda}_{2N+2}}{(N+1)!^2} |\Phi|^{2(N+1)} \; , \qquad
  \end{eqnarray}
then the resulting system will have a bound state of $(N+1)$-particles. Using normal ordering and canonical quantisation we can show through the Heisenberg equation of motion that finding such a bound state is equivalent to determining whether the $(N+1)$-body wavefunction $\Psi(\vec{x}_{1},\ldots,\vec{x}_{N+1})$ satisfies
  \begin{eqnarray}
     \sum_{i=1}^{N} \left( (-\Delta_{i})^{N} + \frac{\tilde{\lambda}_{2N+2}  \prod_{i \neq j} \delta^{(2)}(\vec{x}_{i} - \vec{x}_{j})}{N! (N+1)!} + |E| \right) \Psi = 0
     \nonumber
  \end{eqnarray}
for some energy $-|E|$ where $\Delta_{i}$ is the Laplacian for the position vector $\vec{x}_{i}$. Investigating these theories is beyond the remit of this paper but these ``scalar lumps'' represent rather novel phenomena worthy of independent consideration.}

\section{Quantum charges and Ward identities}

{\ From the conserved currents \eqref{Eq:H}, \eqref{Eq:Pi} and \eqref{Eq:U(1)charge} it is possible to define the usual conserved charges $\hat{H}$, $\hat{Q}$, $\hat{P}_{i}$ and $\hat{M}_{ij}$ (the angular momentum operator). However, in the scale invariant theory we can additionally define a conserved scaling charge
  \begin{eqnarray}
    \hat{D} = \int d^{2} x \; S^{t} = 4 t \hat{H} + \int d^{2}x \; \vec{r} \cdot \vec{\mathcal{P}} \; . 
  \end{eqnarray}
The addition of this conserved charge makes the scale invariant field theory a representation of the Lifshitz algebra. In particular the scaling weights of the other conserved charges are
  \begin{eqnarray}
    \left[ \hat{D}, \hat{H} \right] = 4 i \hat{H} \; ,  &\qquad& \left[ \hat{D}, \hat{Q} \right] = 0 \; , \\ 
    \left[ \hat{D}, \hat{P}_{i} \right] = i \hat{P}_{i} \; , &\qquad& \left[ \hat{D}, \hat{M}_{ij} \right] = 0 \; .
  \end{eqnarray}
As in a conformal theory given an operator $\hat{O}$ with weight $\Delta_{O}$, i.e.
  \begin{eqnarray}
    \left[\hat{D}, \hat{O} \right] = i \Delta_{O} \hat{O}
  \end{eqnarray}
we can use $\hat{P}_{i}$ to create a new operator with higher weight. In particular,
  \begin{eqnarray}
    \left[ \hat{D}, \left[ \hat{P}_{i}, \hat{O} \right] \right] = i \left( \Delta_{O} + 1 \right) \left[ \hat{P}_{i}, \hat{O} \right] \; . 
  \end{eqnarray}
However there is no scaling weight lowering operator in the Lifshitz algebra, which in the conformal algebra would allow the definition of a new operator such that
  \begin{eqnarray}
   \left[ \hat{D}, \left[ \hat{K}_{i}, \hat{O} \right] \right] = i \left( \Delta_{O} - 1 \right) \left[ \hat{K}_{i}, \hat{O} \right] \; , 
  \end{eqnarray}
where $\hat{K}_{i}$ is the special conformal generator. We can certainly find a perturbative unitarity bound for our Lifshitz scalar following \cite{Fujimori:2015mea}, which would give an estimate for the minimal weight of operators, but we have no way of reaching the bottom of the tower in a methodical manner given some arbitrary initial operator.}

{\ With the $\beta$-functions we developed previously we can also discuss the behaviour of the scale Ward identity \eqref{Eq:tracerelation} at the quantum level. From our analysis of the scattering amplitudes we know that unless $\zeta$ is zero the theory will not have scale invariance. At one loop level, it can be argued \cite{PhysRevD.2.1541} that the following operator identity holds
  \begin{eqnarray}
   4 \hat{T}\indices{_{t}^{t}} + \hat{T}\indices{_{i}^{i}} &=& \beta_{\lambda}(\mu) | \hat{\Phi} |^2 | \partial \hat{\Phi} |^2 \; . \qquad
  \end{eqnarray}
Therefore, with our choice of renormalisation scheme and subsequent definition of the couplings, the $\beta$-function of the $\lambda$ coupling parameterises the failure of scale invariance.}

\section{Discussion}

{\ We have constructed what is arguably the simplest interacting quantum field theory with Lifshitz scale invariance. This is a theory of interacting non-relativistic bosons living in $(2+1)$-dimensions. It has an exactly marginal deformation about the Gaussian fixed point that yields a line of perturbatively scale invariant theories. Unusually this marginal interaction has a derivative coupling. We have also claimed, given our results on the sextic interaction, that in a particle-hole asymmetric, charged scalar theory with dynamic critical exponent $2N$ there is a bound state of $(N+1)$ particles.}

{\ Given such a simple theory it is a natural question to ask whether additional field content will preserve the scale symmetry or break it. One direction to proceed is to supersymmetrize our model (which may have additional benefits such as making certain observables protected under dialing the coupling). This is readily achieved for our classically scale invariant theory by adding a suitable kinetic term for singlet fermions \cite{PhysRevD.42.3500} and modifying the interaction terms to be
  \begin{eqnarray}
    \label{Eq:SUSYInteractionHamiltonian}
    &\;& \lambda \left( |\Phi|^2 \|\vec{\partial} \Phi\|^2 + |\Psi|^2 \|\vec{\partial} \Phi\|^2 + |\Phi|^2 \|\vec{\partial} \Psi\|^2 
	  + |\Psi|^2 \| \vec{\partial} \Psi \|^2 \right) \nonumber \\
    &\;& + \frac{\zeta^{*}}{2} \left( |\Phi|^2 \left( \Phi^{*} \vec{\partial}^2 \Phi + \Psi^{*} \vec{\partial}^2 \Psi \right) + |\Psi|^2 \Phi^{*} \vec{\partial}^2 \Phi
			      \right) + \mathrm{c.c}  \nonumber \\
    &\;& + \frac{\tilde{\lambda}}{3!^2} \left( | \Phi |^2 + 3 |\Psi|^2 \right) | \Phi |^4
  \end{eqnarray}
where $\Psi$ is the a complex, scalar Grassman field. The system now has $\mathcal{N}=2$ ``kinematical'' supersymmetry \cite{Leblanc:1992wu}
  \begin{eqnarray}
    \delta \Phi = \eta^{*} \Psi \; , \qquad \delta \Psi = - \eta \Phi \; ,
  \end{eqnarray}
where $\eta$ is a complex Grassman parameter. The corresponding nilpotent supercharge, $\hat{q}$, 
  \begin{eqnarray}
   \hat{q} &=& \int d^{2}x \; \hat{\Phi}^{\dagger} \hat{\Psi} \; , 
  \end{eqnarray}
satisfies 
  \begin{eqnarray}
    \left\{ \hat{q}, \hat{q}^{\dagger} \right\} = \hat{N}_{\Phi} + \hat{N}_{\Psi} \; , \qquad 
  \end{eqnarray}
where $\hat{N}_{\Phi}$ and $\hat{N}_{\Psi}$ are the conserved $U(1)$ number charges for the boson and the fermion respectively and we have imposed canonical commutation relations. Whether this theory is scale invariant or not requires additional analysis, in particular we expect the quartic interaction in $\Psi$ to be divergent as fermionic statistics mean that the vertex depends on the relative momentum of the particles, not the total momentum, which need not be conserved across a vertex (unlike $\vec{P}^2$).}

{\ A second type of supercharge, the ``dynamical'' supercharge \cite{Leblanc:1992wu}, exists for the free version of our Lifshitz model in analogy with the Schr\"{o}dinger case (see \cite{Nakayama:2008qz,Nakayama:2009ku,Xue:2010ih,Tong:2015xaa,Chapman:2015wha,Meyer2017} for example). In this latter situation the nilpotent supercharges $\hat{Q}$ satisfy
  \begin{eqnarray}
    \left\{ \hat{Q}, \hat{Q}^{\dagger} \right\} = \hat{H} \; .
  \end{eqnarray}
These supercharges are precisely those discussed in \cite{Brattan:2018sgc} when one takes a vanishing external potential. In the ``single-particle'' case of that paper it was observed that, independent of the potential, scale invariance can be preserved for suitable choices of the self-adjoint extension. However, in the case under consideration the interaction term of \eqref{Eq:ScaleInvariantAction} is not invariant under dynamical supersymmetry. Such a modified action, with an additional Chern-Simons gauge field, does exist and will be discussed elsewhere.}

{\ The work of \cite{Brattan:2018sgc} suggests introducing spatially extended potential terms might be interesting. A novel way to achieve this is to couple to a Chern-Simon's gauge field. It is well-known that Chern-Simon's terms preserve classical scale invariance (and supersymmetry) and their consequences for the Schr\"{o}dinger theory are well understood \cite{PhysRevD.42.3500,Bergman:1993kq}. Indeed, perhaps promoting the subsequent $U(1)$ gauge symmetry to $U(N)$ could produce interesting theories where scale invariance is preserved. An additional complication to the Schr\"{o}dinger case arises in that the minimial gauging of a kinetic term which is quartic (or higher) in spatial derivatives is inherently ambiguous. This ambiguity can be characterised by additional terms in the action beyond the na\"{i}ve gauging of the kinetic term to $|D^{2} \Phi|^2$, such as $|\Phi|^2 F^2$.}

\vspace{1cm}

\begin{acknowledgments}
{\ The work of DB was supported by key grants from the NSF of China with grant numbers:~11235010 and 11775212. Currently DB is supported by an INFN fellowship from competition n.~19292/2017. DB would like to thank Brian Skinner for discussions on excitons and physical systems having quartic dispersion relations. DB would also like to thank Oren Bergman for suggesting this line of research, Carlos Hoyos for a discussion on SEM tensor complex improvement in Lifshitz theories and Omrie Ovdat for comments on the draft.}
\end{acknowledgments}


\bibliography{references}

\clearpage


\section*{Supplementary material}

# The curious behaviour of the $(2+1)$-dimensional Lifshitz scalar
# Supplementary materials

Daniel K. Brattan*

*Interdisciplinary Center for Theoretical Study, University of Science
and Technology of China, 96 Jinzhai Road, Hefei, Anhui, 230026 PRC.*

(Dated: August 2, 2018)

**Supplementary Note 1 : Evaluating $I_1(X)$, $I_2(X)$ and $I_3(X)$**

We will need to evaluate integrals of the form

$$\int_{\theta=0}^{2\pi} d\theta \frac{\cos^n(\theta)}{1 - (X^2 - 2Xx\cos(\theta) + x^2)^2 - x^4 + i\epsilon} \tag{S1}$$

where $n = 0, 1, 2$. We define

$$\alpha_\pm = \frac{x}{2X} + \frac{X}{2x} \pm \frac{1}{2xX}\sqrt{1 - x^4 + i\epsilon}\,, \tag{S2}$$

use the substitution $\cos(\theta) = (1-z^2)/(1+z^2)$ and extract a factor from the denominator to find:

$$-\frac{1}{X^2 x^2 (1-\alpha_+)(1-\alpha_-)} \int_{z=0}^{\infty} dz \, \frac{(1-z^2)^n}{(1+z^2)^{n-1}\left(1 + \frac{1+\alpha_+}{1-\alpha_+}z^2\right)\left(1 + \frac{1+\alpha_-}{1-\alpha_-}z^2\right)}\,. \tag{S3}$$

For $n = 0, 1, 2$ this expression can be evaluated in terms of elementary integrals. Turning now to the precise expressions used in the text we are required to evaluate

$$I_1(X) = \int^{\tilde{\Lambda}} \frac{dxd\theta}{(2\pi)^2} \frac{x}{1 - (X^2 - 2Xx\cos(\theta) + x^2)^2 - x^4 + i\epsilon}\,, \tag{S4}$$

$$I_2(X) = \int^{\tilde{\Lambda}} \frac{dxd\theta}{(2\pi)^2} \frac{x\left(X^2 - 2Xx\cos(\theta) + 2x^2\right)}{1 - (X^2 - 2Xx\cos(\theta) + x^2)^2 - x^4 + i\epsilon}\,, \tag{S5}$$

$$I_3(X) = \int^{\tilde{\Lambda}} \frac{dxd\theta}{(2\pi)^2} \frac{X^2\left(X - 2x\cos(\theta)\right)^2}{1 - (X^2 - 2Xx\cos(\theta) + x^2)^2 - x^4 + i\epsilon}\,. \tag{S6}$$

After performing the integral over $\theta$ using (S3) we are left with an integral over $x$ which must be done numerically. We use Mathematica's numerical limit package to evaluate these integrals for a range of $X$ with $\epsilon = 10^{-5}$. The results for (S5) and (S6) are given in figs. S1 while (S4) is displayed in figure 3 of the main text.

**Supplementary Note 2 : The stress-energy-momentum tensor complex and its improvement**

The components of the canonical SEM tensor complex are

---

* danny.brattan@gmail.com

$$T_t{}^t = \mathcal{H} = \frac{1}{2m}|\boldsymbol{\partial}^2\Phi|^2 + \lambda|\Phi|^2|\boldsymbol{\partial}\Phi|^2 + \frac{|\Phi|^2}{2}\left(\zeta\Phi^*\boldsymbol{\partial}^2\Phi + (\zeta\boldsymbol{\partial}^2\Phi)^*\Phi\right) + \frac{\tilde{\lambda}}{3!^2}|\Phi|^6 \ , \tag{S7}$$

$$T_i{}^t = \mathcal{P}_i = -\text{Im}\left[\Phi^*\partial_i\Phi\right] \ , \tag{S8}$$

$$\begin{aligned}T_t{}^i = &-\frac{1}{2m}\left(\boldsymbol{\partial}^2\Phi^*\partial_t\partial^i\Phi + \partial_t\partial^i\Phi^*\boldsymbol{\partial}^2\Phi\right) + \frac{1}{2m}\partial^i\boldsymbol{\partial}^2\Phi^*\partial_t\Phi + \frac{1}{2m}\partial_t\Phi^*\partial^i\boldsymbol{\partial}^2\Phi \\ &-\lambda|\Phi|^2\left(\partial^i\Phi^*\partial_t\Phi + \partial_t\Phi^*\partial^i\Phi\right) + 2|\Phi|^2\left(\zeta\partial^i\Phi^*\partial_t\Phi + (\zeta\partial_t\Phi)^*\partial^i\Phi\right) \\ &+\zeta(\Phi^*)^2\partial_t\Phi\partial^i\Phi + \tilde{\lambda}^*\partial_t\Phi^*\partial^i\Phi^*\Phi^2 - \zeta|\Phi|^2\Phi^*\partial_t\partial^i\Phi - \tilde{\lambda}^*|\Phi|^2\partial_t\partial^i\Phi^*\Phi \ ,\end{aligned} \tag{S9}$$

$$\begin{aligned}T_j{}^i = &-\frac{1}{2m}\left(\boldsymbol{\partial}^2\Phi^*\partial_j\partial^i\Phi + \partial_j\partial^i\Phi^*\boldsymbol{\partial}^2\Phi\right) + \frac{1}{2m}\left(\partial_j\Phi^*\partial^i\boldsymbol{\partial}^2\Phi + \partial^i\boldsymbol{\partial}^2\Phi^*\partial_j\Phi\right) - \frac{i}{2}\Phi^*\partial_t\Phi\delta_j{}^i + \frac{i}{2}\partial_t\Phi^*\Phi\delta_j{}^i \\ &+\mathcal{H}\delta_j{}^i - \lambda|\Phi|^2\left(\partial^i\Phi^*\partial_j\Phi + \partial_j\Phi^*\partial^i\Phi\right) + \zeta(\Phi^*)^2\partial_j\Phi\partial^i\Phi + \tilde{\lambda}^*\Phi^2\partial_j\Phi^*\partial^i\Phi^* \\ &+|\Phi|^2\left(2\zeta\partial^i\Phi^*\partial_j\Phi + 2\zeta^*\partial_j\Phi^*\partial^i\Phi - \zeta\Phi^*\partial_j\partial^i\Phi - \tilde{\lambda}^*\partial_j\partial^i\Phi^*\Phi\right) \ ,\end{aligned} \tag{S10}$$

We have explicitly checked that the SEM tensor complex conservation equations are satisfied when the equations of motion for the fields $\Phi$ and $\Phi^*$ are satisfied. However, the purely spatial component $T_{ij}$ does not satisfy the naïve rotation Ward identity and the trace relation has a non-vanishing virial current term. Consequently the SEM tensor complex requires improvement [S1, S2].

The canonical SEM tensor $T_\nu{}^\mu$ of a relativistic conformal field theory satisfies the conservation equation

$$\partial_\mu T_\nu{}^\mu = 0 \tag{S11}$$

and a scaling Ward identity of the form

$$T_\nu{}^\mu = \partial_\mu K^\mu \ . \tag{S12}$$

It can only be improved to give $K^\mu = 0$ if $K^\mu = \partial^\mu V$ for some $V$ [S1]. The improvement term need not have this form for a non-relativistic SEM tensor complex as space and time are treated differently [S2]. Given that only the $\lambda$ coupling leads to a scale invariant field theory at the quantum level we only record an improved SEM tensor complex in this case:

$$T_t{}^t = \frac{1}{2m}|\boldsymbol{\partial}^2\Phi|^2 + \lambda|\Phi|^2|\boldsymbol{\partial}\Phi|^2 + \left[\frac{i}{2}\left(\Phi^*\partial_t\Phi - \partial_t\Phi^*\Phi\right) - \frac{3}{2m^2}\partial_i\partial_j\Phi^*\partial^i\partial^j\Phi\right] \ , \tag{S13}$$

$$T_i{}^t = -\text{Im}\left[\Phi^*\partial_i\Phi\right] \ , \tag{S14}$$

$$\begin{aligned}T_t{}^i = &-\frac{1}{2m}\left(\boldsymbol{\partial}^2\Phi^*\partial_t\partial^i\Phi + \partial_t\partial^i\Phi^*\boldsymbol{\partial}^2\Phi\right) + \frac{1}{2m}\partial^i\boldsymbol{\partial}^2\Phi^*\partial_t\Phi + \frac{1}{2m}\partial_t\Phi^*\partial^i\boldsymbol{\partial}^2\Phi \\ &-\lambda|\Phi|^2\left(\partial^i\Phi^*\partial_t\Phi + \partial_t\Phi^*\partial^i\Phi\right) \\ &-\frac{i}{8m^2}\left[\left(\partial^i\boldsymbol{\partial}^6\Phi^*\Phi - \Phi^*\partial^i\boldsymbol{\partial}^6\Phi\right) + \left(\partial^i\partial_k\boldsymbol{\partial}^4\Phi^*\partial^k\Phi - \partial^k\Phi^*\partial^i\partial_k\boldsymbol{\partial}^4\Phi\right)\right. \\ &\left.+ 5\left(\partial^i\partial_k\partial_l\boldsymbol{\partial}^2\Phi^*\partial^k\partial^l\Phi - \partial^k\partial^l\Phi^*\partial^i\partial_k\partial_l\boldsymbol{\partial}^2\Phi\right) + 5\left(\partial^i\partial_k\partial_l\Phi^*\partial^k\partial^l\boldsymbol{\partial}^2\Phi - \partial^k\partial^l\boldsymbol{\partial}^2\Phi^*\partial^i\partial_k\partial_l\Phi\right)\right] \ ,\end{aligned} \tag{S15}$$

$$\begin{aligned}T_j{}^i = &-\frac{1}{2m}\left(\boldsymbol{\partial}^2\Phi^*\partial_j\partial^i\Phi + \partial_j\partial^i\Phi^*\boldsymbol{\partial}^2\Phi\right) + \frac{1}{2m}\left(\partial_j\Phi^*\partial^i\boldsymbol{\partial}^2\Phi + \partial^i\boldsymbol{\partial}^2\Phi^*\partial_j\Phi\right) \\ &-\frac{i}{2}\Phi^*\partial_t\Phi\delta_j{}^i + \frac{i}{2}\partial_t\Phi^*\Phi\delta_j{}^i + \mathcal{H}\delta_j{}^i - \lambda|\Phi(x)|^2\left(\partial^i\Phi^*(x)\partial_j\Phi(x) + \partial_j\Phi^*(x)\partial^i\Phi(x)\right) \\ &+\left[\frac{3}{2m}\left(\boldsymbol{\partial}^4\Phi^*\Phi + \Phi^*\boldsymbol{\partial}^4\Phi\right)\delta^i_j - \frac{1}{2m}\left(\partial_k\boldsymbol{\partial}^2\Phi^*\partial^k\Phi + \partial_k\Phi^*\partial^k\boldsymbol{\partial}^2\Phi\right)\delta^i_j - \frac{1}{m}\boldsymbol{\partial}^2\Phi^*\boldsymbol{\partial}^2\Phi\delta^i_j\right. \\ &-\frac{3}{m}\partial_k\partial_l\Phi^*\partial^k\partial^l\Phi\delta^i_j - \frac{2}{m}\left(\partial_j\partial^i\boldsymbol{\partial}^2\Phi^*\Phi + \Phi^*\partial_j\partial^i\boldsymbol{\partial}^2\Phi\right) + \frac{2}{m}\left(\partial_j\partial^i\Phi^*\boldsymbol{\partial}^2\Phi + \boldsymbol{\partial}^2\Phi^*\partial_j\partial^i\Phi\right) \\ &-\frac{1}{m}\left(\partial_i\boldsymbol{\partial}^2\Phi^*\partial_j\Phi + \partial_j\boldsymbol{\partial}^2\Phi^*\partial_i\Phi + \partial_i\Phi^*\partial_j\boldsymbol{\partial}^2\Phi + \partial_j\Phi^*\partial_i\boldsymbol{\partial}^2\Phi\right) \\ &\left.+\frac{3}{m}\left(\partial_i\partial_j\partial_k\Phi^*\partial^k\Phi + \partial^k\Phi^*\partial_i\partial_j\partial_k\Phi\right) - \frac{\lambda}{2}\left(\partial_i\partial_j|\Phi|^4 - \delta_{ij}\partial^2|\Phi|^4\right)\right] \ .\end{aligned} \tag{S16}$$

The majority of additional terms are required to make the canonical SEM tensor complex of the free theory satisfy the rotation and scaling Ward identities. Only a small number of improvement terms involve $\lambda$. The reader can verify

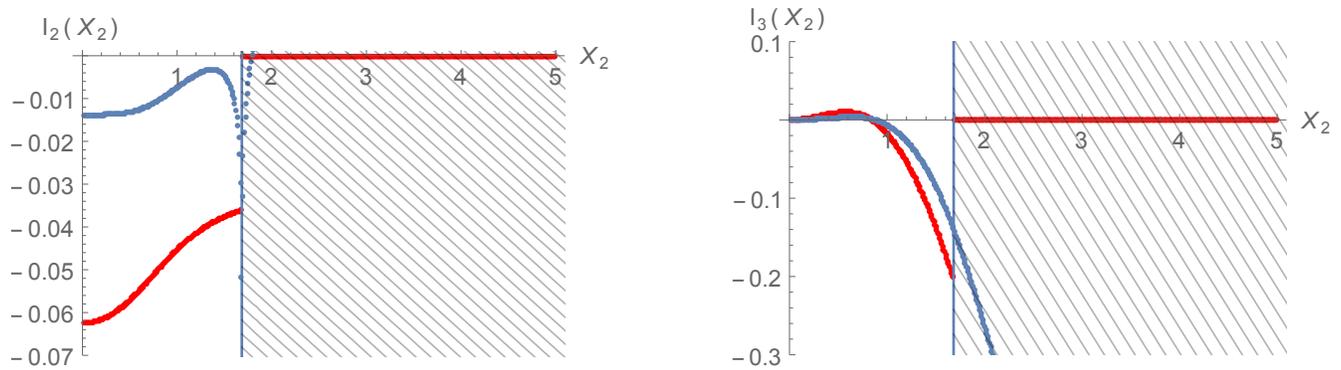

FIG. S1. Plots of the real (blue dots) and imaginary (red dots that are approximately zero in the shaded region) parts of the one-loop contributions represented by integrals (S5) and (S6) respectively. We plot the value of this contribution against the kinematic variable $X$ with $\epsilon = 10^{-5}$. The shaded region has a kinematic parameter greater than the maximum on-shell value.

that it satisfies the trace relation without a virial current.

---



\end{document}